\documentstyle[aps,prl,epsfig,multicol,times]{revtex}
\makeatletter
\newlength\abovecaptionskip \newlength\belowcaptionskip
\def\@makecaption#1#2{%
 \vskip\abovecaptionskip \sbox\@tempboxa{#1: #2}%
 \ifdim \wd\@tempboxa >\hsize #1: #2\par \else \global \@minipagefalse
 \hb@xt@\hsize{\hfil\box\@tempboxa\hfil}%
 \fi \vskip\belowcaptionskip} \makeatother

\newcommand{\var}{\langle \epsilon^2 \rangle}
\newcommand{\mmv}{\langle m^2 \rangle}
\newcommand{\mvar}{\langle m^2 \rangle}
\newcommand{\mav}{\langle m \rangle}
\newcommand{\eav}{\langle \eps \rangle}
\newcommand{\eps}{\epsilon}
\newcommand{\ve}{\vert}
\newcommand{\fr}{\frac}
\newcommand{\ran}{\rangle}
\newcommand{\lan}{\langle}
\newcommand{\beq}{\begin{equation}}
\newcommand{\eeq}{\end{equation}}

\begin{document}            

\title{Scaling of the localization length in linear electronic and 
vibrational systems with long-range correlated disorder}
\author{Stefanie  Russ\\
Institut f\"ur Theoretische Physik III, Universit\"at Giessen,
D-35392 Giessen, Germany \\
\date{\today}
}
 
\maketitle

\begin{multicols}{2}[%
\begin{abstract} 
The localization lengths $\lambda$ of one-dimensional disordered systems are 
studied for electronic wavefunctions in the Anderson model and for vibrational 
states. In the first case, the site energies $\eps$ and in the second case, 
the fluctuations of the vibrating masses $m$ at distance $\ell$ from each 
other are long-range correlated and described by a correlation 
function $C(\ell)\sim\ell^{-\gamma}$ with $0<\gamma\le 1$. 
In the Anderson model, we focus on a scaling theory that
applies close to the band edge, i.e. at energies $E$ close to $2$. 
We show that $\lambda$ can be written as
$\lambda=\lambda_0 f_\gamma(x)$, with $\lambda_0=\var^{-1/(4-\gamma)}\sim
\lambda(E=2,\var)$, $x=\lambda_0^2 (2-E)$ and the scaling 
function $f_\gamma(x)=\rm{const}$ for $x\ll 1$ and $f_\gamma(x)\sim
x^{(3-\gamma)/2}$ for $x\gg 1$.
Mapping the Anderson model onto the vibrational problem, we derive the 
vibrational localization lengths for small eigenfrequencies $\omega$,
$\lambda\sim \mav^{(3-\gamma)/2}\,\mvar^{-1}\,\omega^{-(1+\gamma)}$, where
$\mav$ is the mean mass and $\mvar$ the variance of the masses. 
This implies that, unexpectateley, at small $\omega$, $\lambda$ is 
larger for uncorrelated than for correlated chains. 
\end{abstract}
]

\section{Introduction and Models}

A large amount of work has been done in the past decades to understand 
localization behavior in randomly disordered chains. Most work has been
concentrated on uncorrelated \cite{kramer} and short-range correlated 
disorder ( see \cite{izraelev} and references therein). 
In this work, the effect of power-law long-range correlated disorder 
on the localization properties of linear electronic and vibrational systems 
is discussed. 

First we consider the Schr\"odinger equation of a
single-particle electronic wavefunction in tight binding 
approximation (Anderson model) \cite{anderson}, 
\begin{equation} \label{anderson}
\psi_{n+1} + \psi_{n-1} - 2 \psi_n = -(2-E+\eps_n) \psi_n.
\end{equation}
Here $E$ is the energy and $\ve\psi_n\ve^2$ is the probability to find an
electron at site $n$, with $n\in\{1,N\}$. The $\eps_n$ are the site potentials
and are random numbers, taken from an interval of width $\Delta$,
$\eps_n\in[-\Delta/2,\Delta/2]$ (diagonal disorder).
$\var \equiv (1/N)\sum_{n=1}^N \eps_n^2$ is their 
variance and the average value $\langle\eps\rangle\equiv (1/N)\sum_{n=1}^N 
\eps_n$ is zero. The term $-2\psi_n$ on both sides is introduced 
for technical reasons.

Second, we investigate the related problem of the localization length of 
vibrational states in disordered chains.
In this case, the wavefunction $\psi_n$ of Eq.~(\ref{anderson}) is
substituted by the vibrational amplitudes $u_n$ of $N$ particles at sites $n$,
coupled by unit spring constants  between them. This problem is described by
the time-independent eigenvalue equation
with the vibrational amplitudes $u_{n}$ and the eigenfrequencies $\omega$: 
\begin{equation} \label{schwing}
u_{n+1} + u_{n-1} - 2 u_{n} = -m_n \omega^2 u_{n}
\end{equation}
Contrary to the $\eps_n$ of Eq.~(\ref{anderson}), all masses $m_n$ are positive 
and taken from an interval $m_n\in[\mav-\Delta/2,\mav+\Delta/2]$, where
$\langle m \rangle \equiv (1/N)\sum_{n=1}^N m_n> 0$ is their mean value
and $\mmv\equiv (1/N)\sum_{n=1}^N (m_n-\lan m\ran)^2$ the variance. 

The localization length $\lambda$ in the Anderson case is defined by the 
exponential decay of the wavefunctions $\psi_n$, 
%\beq \label{loclength}
$\lim_{n\to\infty}\ve\psi_n/\psi_0\ve = \exp[-n/\lambda]$.
%\eeq
In the vibrational problem, $\psi_n/\psi_0$ is substituted by $u_n/u_0$, which
decays accordingly. 

In the standard Anderson model with uncorrelated diagonal disorder
it has been recognized since long that in $d=1$ all states are exponentially 
localized \cite{kramer,borland,halperin}, i.e., that $\lambda$
approaches a constant for large $N$. 
A weak disorder expansion yields $\lambda =
\var^{-1/3}f((2-E)/\var^{2/3})$ close to the band edge \cite{gid,dergard}, 
while at the band center, a Green's function technique yields
$\lambda\sim \var^{-1}$ \cite{kappus}. 
In some distance from the band center, the diagonal elements of the Green's
functions can be developed by a second-order perturbation theory, yielding 
$\lambda(E) \sim (4-E^2)/\var$ \cite{thouless}. 

In the related vibrational problem with uncorrelated random masses,
it is well accepted that all states in $d=1$ are localized, except for the case
$\omega=0$, where the disorder term $m_n \omega^2$ disappears.
Transfer matrix calculations in $d=1$ yield
$\lambda\sim\omega^{-2}$ for small $\omega$ \cite{azbel}.

Here, we concentrate
on long-range correlated disorder, characterized by a correlation function 
$C(\ell)\sim\ell^{-\gamma}$. In the Anderson case, $C(\ell)$ describes the 
correlations between site energies $\eps_n$ and $\eps_{n+\ell}$ at sites $n$ and
$n+\ell$,
\begin{equation} \label{corr1}
C(\ell) = \lim_{N\to\infty}\frac{1}{N} \sum_{n=1}^N \eps_n \eps_{n+\ell} 
\equiv\lan\eps_n \eps_{n+\ell}\ran\sim \ell^{-\gamma}
\end{equation}
with the correlation exponent $\gamma$, $0<\gamma\le 1$. In the vibrational 
problem, the mass fluctuations are long-range correlated and described
by $C(\ell)=\lan\tilde m_n\tilde m_{n+\ell}\ran\equiv
\ell^{-\gamma}$ with $\tilde m_n\equiv m_n-\langle m \rangle$.
The case $\gamma\ge 1$ describes only short-range correlated potentials. 
From random walk theory, we expect that series with
$\gamma \ge 1$ fall into the same universality class as uncorrelated 
potentials \cite{Bunde2,prb}.

For the one-dimensional Anderson model of Eq.~(\ref{anderson}) with long-range
correlated potentials, a scaling form for the localization length $\lambda(E,\var)$
close to the band edge $E=2$ has been developed recently and supported by
preliminary numerical simulations \cite{philmag}. 
For the vibrational problem, the localization length in the presence of 
correlations has not been adressed yet.
In this paper, we determine explicitely the scaling function 
for the Anderson system and derive its asymptotic form by scaling arguments 
(section II). This leads also to an explicit expression for
$\lambda$ at intermediate energies in some distance from the band edge 
(cf. Eq.(\ref{acomplet})).
In section III, we map the Anderson model onto the vibrational problem and 
show that for small $\omega$, $\lambda(\omega) \sim\omega^{-(1+\gamma)}$ 
which agrees with the 
result $\lambda \sim\omega^{-2}$ of \cite{azbel} for random uncorrelated 
potentials, if we describe them by $\gamma=1$. Accordingly, contrary to the 
expectations, $\lambda$ is decreased by the presence of long-range 
correlations for $\omega\to 0$.

\section{The long-range correlated Anderson model}

In a previous work \cite{philmag}, it was shown by a space-renormalization 
procedure that  at the band edge, $E=2$, the localization length $\lambda$ 
scales as
\begin{equation}\label{skal1}
\lambda(E=2,\var)\sim\var^{-1/(4-\gamma)}\equiv \lambda_0, 
\end{equation}
For $E<2$, but still in the neighborhood of $2$, $\lambda$ depends on both, 
$\lambda_0$ and $2-E$ and can be written as: 
\begin{eqnarray} 
\lambda(E,\var) &=& \lambda_{\rm{0}} \, \,f_\gamma(x), \label{skalfkt}\\
\mbox{where}\quad x = \lambda_0^2 \,(2-E) &=& (2-E)\var^{-2 /(4-\gamma)}
\label{xarg}
\end{eqnarray}
and $f_\gamma(x)$ is a correction function that depends on $\gamma$ and 
approaches a constant for $x\to 0$.
For $\gamma = 1$, one recovers the exponent $1/(4-\gamma) = 1/3$, in accordance
with the result of \cite{dergard} for uncorrelated potentials. 

We are interested in the behavior of $f_\gamma(x)$ in the limits of large and
small values of $x$.
For uncorrelated potentials $\eps_n$, we know already that 
$f_{\rm{uncorr}}(x) \to \rm{const}$ for $x\ll 1$ \cite{dergard} and 
$f_{\rm{uncorr}}(x) \to x$ for $x \gg 1$ \cite{kappus,thouless}.
We therefore assume a power-law behavior 
\beq
f_\gamma(x) \sim x^\alpha
\eeq
for correlated potentials, where $\alpha$ depends on $\gamma$.

To derive the exponent $\alpha$, it is helpful to
realize that $1/(2-E)$ is proportional to the square of the wavelength 
$\Lambda$ of the electronic wavefunction ($\eav=0$),
\begin{equation} \label{wavelength}
\Lambda^2\sim 1/(2-E).
\end{equation}
This can be seen most easily in an ordered 
chain, where all $\eps_n=0$ and Eq.~(\ref{anderson})
reduces to $\psi''(x)/\psi(x) = -(2-E)$. The
solution is a harmonic function with the wavelength $\Lambda$.
In the case of disorder,
we have calculated many states of correlated and uncorrelated chains in 
different disorder intervals $\Delta$ by the Lanczos algorithm and 
have verified by Fourier transformation that relation (\ref{wavelength})
still holds. So, $\Lambda$ describes the oscillating part of the wavefunction, 
while $\lambda$  describes the exponential decay of the envelope. A
similar relation for the wavelength has been found for vibrations
in percolation clusters in $d=2$ and $d=3$ in \cite{kantelhardt98}.

Next, we look at a wavefunction with a given wavelength $\Lambda$ and 
discuss, how $\lambda$ is influenced by a change of the variance of the
potentials $\var$, i.e. by a change of $\lambda_0$.
We discuss the cases $x\ll 1$ and $x \gg 1$ separately, which correspond
to $\Lambda \gg\lambda_{\rm{0}}$ and $\Lambda \ll\lambda_{\rm{0}}$, 
respectively. We have to keep in mind that 
$\Lambda$ is solely determined by $E$, whereas $\lambda_{\rm{0}}$ 
is determined by $\var$ and $\gamma$. 

(i) For $\Lambda \gg\lambda_{\rm{0}}$, close to the band edge, 
the wave amplitude decays within the first wavelength $\Lambda$. Accordingly,
the wavefunction is independent of $\Lambda$ and would not change for
$\Lambda\to\infty$, i.e., by approaching the band edge, where $\lambda \sim 
\lambda_0\equiv\var^{-1/(4-\gamma)}$. 
Hence we expect for all $\gamma$ that the scaling function 
$f_\gamma(x)=\lambda/\lambda_0$ approaches a constant for small $x$. 

(ii) In the opposite case, $\Lambda \ll \lambda_{\rm{0}}$, 
the wave amplitude performs many
oscillations before the envelope function decays completely.
It is reasonable to assume that for the dependence of $\lambda$ on
the potential landscape, long-range correlations 
are not relevant, when the wavefunction oscillates rapidly. In this
case, the term $\eps_n\psi_n$ on the right-hand side of Eq.~(\ref{anderson})
oscillates rapidly for both, correlated and uncorrelated potentials. 
So, in the case $\Lambda \ll \lambda_0$, we expect the same dependence of 
$\lambda$ on $\var$ for correlated and for uncorrelated potentials. 
For the latter, we know that in some distance from the band edge,  
the behavior of $\lambda$ must cross over to the
result of \cite{thouless}, 
%\begin{equation} \label{lv}
$\lambda\sim\var^{-1}$ 
%\end{equation} 
and in view of the preceeding remark, we expect the same behavior also
for correlated potentials.
Inserting this relation into (\ref{skalfkt}) and (\ref{xarg}), we find 
$f_\gamma(x)\sim x^\alpha$, with $\alpha=(3-\gamma)/2$. 
In summary, we find for the asymptotic cases of the scaling function
$f_\gamma(x)$ the power-law behavior
\begin{equation}\label{assym}
f_\gamma(x) \sim x^\alpha = \cases {\mbox{const.} & $x \ll 1$ \cr 
x^{(3-\gamma)/2} & $x \gg 1$ \cr}
\end{equation}

Inserting Eq.~(\ref{assym}) into (\ref{skalfkt}) and (\ref{xarg}), we find 
for the energy dependence in the case $\Lambda \ll \lambda_{\rm{0}}$ with the
additional constraint $2-E\ll 1$
\begin{equation}\label{acomplet}
\lambda(E,\var)\sim \var^{-1}\,(2-E)^{(3-\gamma)/2}.
\end{equation}
Accordingly, we can distinguish between three energy regimes. (a) For
$2-E\ll \var^{-2/(4-\gamma)}$, i.e., very close to the band edge, we have
$\lambda\sim\var^{-1/(4-\gamma)}$, independent of $E$. 
(b) For $1\gg 2-E\gg \var^{-2/(4-\gamma)}$, Eq.~(\ref{acomplet}) applies. 
(c) For still larger values of $2-E$, the scaling
behavior breaks down. 

For growing values of $2-E$, it was obtained that in some distance from the 
band center $\lambda(E) \sim (2-E)(2+E)/\var$ for uncorrelated potentials 
\cite{thouless}.
This expression emerges into Eq.~(\ref{acomplet}), if we set $\gamma=1$,
which yields $\lambda\sim (2-E)/\var$. 
So we can see again that the case of uncorrelated potentials is well described
by $\gamma=1$.

\begin{figure}\centering 
\epsfysize6cm\epsfbox{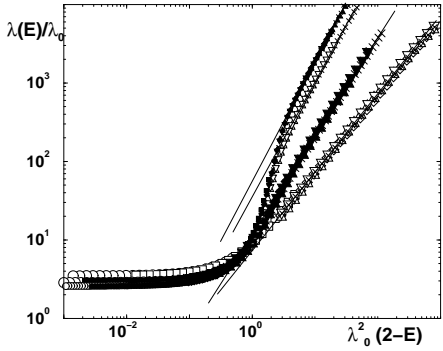} 
\parbox{8cm}{
\caption[]{\small 
Plot of the scaling function $f_\gamma(x)= \lambda(E)/\lambda_0$ versus
$x=\lambda_0^2\,(2-E)$ close to band edge for the 
Anderson model in $d=1$. The curves show (from bottom to
top): uncorrelated disorder (large open symbols), 
correlated site potentials of $\gamma=0.5$ (large filled symbols), $\gamma=0.1$ 
(small open symbols) and $\gamma=0.01$ (small filled symbols).
The symbols stand for $2-E = 10^{-4}$
(circles), $10^{-3}$ (squares), $10^{-2}$ (diamonds), $0.1$ (triangles up),
$0.2$ (triangles down), $0.3$ (x) and $\var$ was varied between $10^{-5}$ and
$10^{-1}$. 
The typical average of $\lambda$ was performed over $1000$ chains of
length $N=2^{20}$. The straight lines of the theoretical slopes
$(3-\gamma)/2$ are guides to the eye.
}
\label{fig:1}}
\end{figure}

In order to test this scaling function, we have generated 
$\nu=1000$ linear chains of size $N = 2^{20}$ 
by the method of Fourier transformation \cite{fourier} and calculated 
the individual localization lengths $\lambda_i$ by the transfer-matrix method 
\cite{kramer,prb,dermechpich}.
To obtain the mean localization length, we performed the typical (logarithmic) 
average 
$\lambda(E,\var)\equiv\lambda_{\rm{typ}}=\exp[\fr{1}{\nu}\sum_{i=1}^\nu
\ln\lambda_i]$.

To obtain the scaling function $f_\gamma(x)$, we plotted
$\lambda/\lambda_0$ versus $x=\lambda^2_{\rm{0}}\, (2-E)$. The results are
shown in Fig.~1 for (from bottom to top) uncorrelated potentials 
(described by $\gamma=1$) and correlated potentials with
$\gamma=0.5,0.1$ and $0.01$. 
Five values of $(2-E)$ between $10^{-4}$ and $0.3$ were calculated while 
$\var$ varied between $\var=10^{-5}$ and $10^{-1}$. The figure supports the 
scaling theory. The data for the different $2-E$ and $\var$ fall onto the 
same scaling functions $f_\gamma(x)$, which show the
expected behavior (cf. Eq.~(\ref{assym})) in the asymptotic cases.
The lines are guides to the eye with the theoretical slopes of 
$\alpha=(3-\gamma)/2$.
A crossover close to $x=1$ separates the cases $x\ll 1$ and $x\gg 1$, i.e. 
$\lambda^2_{\rm{0}} \ll 1/(2-E)$ and $\lambda^2_{\rm{0}} \gg 1/(2-E)$.
In the case $x\ll 1$, we find $\lambda/\lambda_0=\mbox{const}$, i.e. $\alpha=0$.

\section{The long-range correlated vibrational model}

This scaling theory can be generalized to the vibrational problem, described
by Eq.~(\ref{schwing}). %Since the average values of $\eav$ and $\mav$ 
%differ ($\eav=0$ and $\mav> 0$), 
We split $m_n$ into the average part $\mav$ and the
fluctuation part $\tilde m_n$, $m_n = \mav + \tilde m_n$ and can  
map the Anderson problem onto the vibrational problem by
replacing
\begin{equation} \label{map}
2-E \to \omega^2 \mav \qquad\mbox{and}\qquad \eps_n \to \omega^2 \tilde m_n.
\end{equation}
We obtain the wavelength $\Lambda\sim (\mav^{1/2}\,\omega)^{-1}$ 
by inserting (\ref{map}) into (\ref{wavelength}) and the
scaling variable $x$  by inserting (\ref{map}) into (\ref{xarg}), 
yielding (with $\lan\tilde m^2\ran = \mvar$)
\beq \label{skalvib2}
x=\mav\,\omega^{-2\gamma/(4-\gamma)}\mvar\,^{-2/(4-\gamma)}.
\eeq
\begin{figure}\centering 
\epsfysize6cm\epsfbox{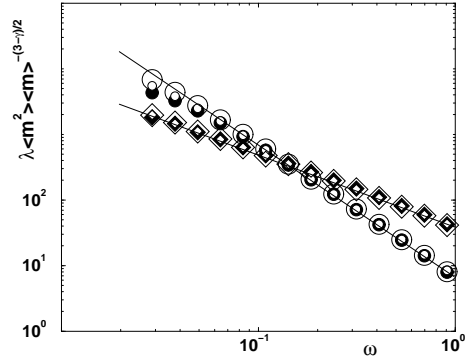} 
\parbox{8cm}{
\caption[]{\small 
The scaled localization length $\lambda\,\mvar \mav^{-(3-\gamma)/2}$
for the vibrational chain is plotted versus the eigenfrequency $\omega$ for  
uncorrelated masses (circles) and correlated masses of $\gamma=0.1$ 
(diamonds) for different $\mav$ and disorder intervals $\Delta$: $\mav=0.5$,
$\Delta=0.99$ (large open symbols), $\mav=0.5$, $\Delta=0.5$ (filled symbols) 
and $\mav=0.25$, $\Delta=0.49$ (small open symbols). The values of $\Delta=0.99$ 
and $0.5$ refer zu $\mvar\approx 0.008$ and $0.002$, respectiveley.
The typical average of $\lambda$ was performed over $1000$ chains of
length $N=2^{23}$. The lines with the slopes $1+\gamma$ are guides to the eye. 
Small deviations from the straight line at very large values of 
$\lambda\approx N/2$ are finite-size effects.
}
\label{fig:2}}
\end{figure}
Since $\mvar<\mav^2$, we can easily verify that for
$\omega^2 < 1/\mav$, the scaling variable $x$ is greater than $1$. Therefore,
contrary to the Anderson problem, we have only one scaling regime $x>1$.
Inserting (\ref{map}) into (\ref{acomplet}),
we find $\lambda$ for the vibrational problem:
\beq \label{dispers}
\lambda \sim 
\mav^{(3-\gamma)/2}\,\mvar^{-1}\,\omega^{-(1+\gamma)}.
\eeq 
This relation is valid for small values of $\omega^2\mav$. 
In our simulations, we found that for e.g. $\mav=0.5$, it holds approximately
for $\omega < 1$.
In order to test Eq.~(\ref{dispers}), we have calculated
the typical average $\lambda_{\rm{typ}}$ of $1000$ chains of length $2^{23}$
for different values of $\gamma$, $\mav$ and $\mvar$ by the transfer-matrix 
method. In Fig.~2, the scaled localization length 
$\lambda\,\mvar\,\mav^{-(3-\gamma)/2}$ is 
plotted versus $\omega$ for $\gamma=0.1$ and for uncorrelated random
chains (described by $\gamma=1$). 
Both, the data collapse for chains of different $\mav$ and $\mvar$ and the
slopes support Eq.~(\ref{dispers}). 
The lines are guides to the eye with the theoretical slopes of $-(1+\gamma)$
and describe the behavior of $\lambda$ very well. 

Two phenomena are worth mentioning here. First, $\lambda$ diverges if 
$\omega\to 0$. The reason is that the disorder term $m_n$ 
in (\ref{schwing}) appears only in combination with the eigenvalue $\omega^2$ 
and therefore disappears for $\omega\to 0$. Second, for
large $\omega$, $\lambda$ is larger
for the correlated than for the uncorrelated chains. However, below some
crossover frequency (that depends on $\gamma$), this behavior changes
and the vibrational excitations of the correlated chains become more localized.
This, at the first glance unexpected behavior can be understood as follows. 

Large $\omega$ correspond to small wavelengths $\Lambda$. Particles $m_n$ in
distances less than $\Lambda/2$ from each other move into the same direction 
and can roughly be considered as an effective hyperparticle $M_j \sim
\sum_{n=1}^\mu m_n$,
where the sum runs over all $\mu$ particles in the region of the size of
$\Lambda/2$. The effective disorder seen by the vibrational excitation 
depends on the
distribution of the $M_j$ in the region, where the wave amplitude is
large. For strongly correlated chains, there exist large regions of masses 
$m_n$ that are either below the average mass $\mav$ or above. So,  
as long as $\Lambda/2$ is smaller than the size of these regions,
the distribution of the $M_j$ is more narrow for correlated than for
uncorrelated chains, which leads to larger $\lambda$.

This behavior changes when the size of $\Lambda/2$ is well above 
the size of the correlated regions. Now, on scales of $\Lambda/2$ the 
system looks much more disordered than before. We know 
from random-walk theory that  for large $\mu$ the variance 
of the $M_j$ scales as $\sim \mu^{2-\gamma}$ \cite{Bunde2}. Hence, 
for large $\mu$ the distribution $M_j$ is broader in correlated than in 
uncorrelated chains and therefore, the disorder seen by the vibrational 
excitation is even larger. Therefore,
we expect smaller $\lambda$ for the correlated chains, when $\omega\to 0$.

The same effect appears also in the Anderson model, where, according to
Eq.~(\ref{acomplet}), $\lambda$ is also smaller for correlated than for
uncorrelated chains in the limit of small values of $2-E$. For systems at the 
band edge, this has already been reported in \cite{philmag}.

\section{Conclusions}

In summary, we have studied the effect of long-range correlated disorder 
on the localization lengths of electron states in the one-dimensional Anderson 
model and of the vibrational states of
harmonic chains. In the Anderson model, the site potentials 
$\eps$ are long-range correlated, while in the vibrational 
problem we have long-range correlations in the
fluctuation part $\tilde m_n$ of the masses around their mean value $\mav$. 

First, we studied the scaling behavior of the localization length $\lambda$ in 
the Anderson model close to the band edge $E=2$. We have shown that
two characteristic lengths occur: the wavelength $\Lambda$ that
describes the oscillating part of the wavefunction and the localization length 
$\lambda_0$ that describes how the envelope of the wavefunction decays at
the band edge. While $\Lambda$ depends on the energy $E$, 
$\lambda_0$ depends on the variance $\var$ 
and the correlation exponent $\gamma$. We developed a scaling theory, with the
scaling variable $x=\lambda_0^2/\Lambda^2$. Using qualitative arguments, we
derived the asymptotic behavior of the scaling function, both, for
$\Lambda\gg\lambda_0$ and $\Lambda\ll\lambda_0$, and thus the
behavior of $\lambda$ in these limits. It was found that 
$\lambda$ is independent of $E$ at very small values of $2-E$. 
For larger distances from the band edge, but $E$ still in the neighborhood of
$2$, it depends on $2-E$ and on $\var$ by power-laws.

Second, we developed an analogous scaling theory for the
vibrational problem by mapping the Anderson equation onto the equations of 
motion of harmonic vibrations. We found that 
$\lambda(\omega)\sim\omega^{-(1+\gamma)}$ in the limit of small frequencies and
for $0<\gamma\le 1$. This relation may constitute a possibility to measure 
correlations by determining the localization length of
vibrations. It implies that $\lambda$ grows faster for uncorrelated than
for correlated chains, when $\omega\to 0$. This is in close analogy to the
behavior of $\lambda$ in the Anderson model at the band edge and
arises from the coupling of correlated masses over very large distances, 
which leads to larger fluctuations than in the case of uncorrelated potentials.
Accordingly, for very small frequencies, correlated chains have smaller
localization lengths than uncorrelated ones.

I gratefully acknowledge valuable discussions
with Armin Bunde, Shlomo Havlin, Jan Kantelhardt and Itzhak Webman and 
financial support from the Minerva foundation.

\end{multicols}

\end{document}